\documentclass[fleqn,twoside]{article}
\usepackage{espcrc2}
\usepackage{graphicx}

\def \ep{\vspace{-1mm}}
\def \det{\mathop{\rm det}}
\def \tr{\mathop{\rm tr}}
\def \ln{\mathop{\rm ln}}

\def \seff{\mathop{S_{\rm eff}}}

\newcommand{\AmS}{{\protect\the\textfont2
  A\kern-.1667em\lower.5ex\hbox{M}\kern-.125emS}}

\title{Comparing the $R$ algorithm and RHMC for staggered fermions}

\author{M. A. Clark, B. Jo\'{o} and A. D. Kennedy\\School of
Physics, The University of Edinburgh, Edinburgh, EH9 3JZ, UK}

\begin{document}
\pagestyle{empty}

\begin{abstract}
The $R$ algorithm is widely used for simulating two flavours of
dynamical staggered fermions.  We give a simple proof that the
algorithm converges to the desired probability distribution to within
$O(\delta\tau^2)$ errors, but show that the relevant expansion
parameter is $(\delta\tau/m)^2$, $m$ being the quark mass.  The
Rational Hybrid Monte Carlo (RHMC) algorithm provides an exact (i.e.,
has no step size errors) alternative for simulating the square root of
the staggered Dirac operator.  We propose using it to test the validity of the
$R$ algorithm for simulations carried out with $\delta\tau\approx m$.
\vspace{0.6pc}
\end{abstract}

\maketitle

\section{Introduction}
 The motivation for this work is the need to decide which lattice
action to use in the future, to allow dynamical fermion simulations
using as light a quark mass as possible.  Recent UKQCD research has
been performed using Wilson fermions, but the computational cost of
producing ensembles is too great for $m_\pi/m_\rho<0.4$.  One
alternative conclusion is to use two flavours of improved staggered
fermions, since the cost appears to scale significantly better for
small masses.

Clearly, it is important to verify that the $R$ algorithm is correct
for small quark masses: this is the subject of this investigation.
The other possible failings of two flavour staggered fermions will not
be addressed here.

\section{Deriving the $R$ Algorithm}

\subsection{The $\Phi$ Algorithm}
We start with the probability distribution for gauge field $U$ and
pseudofermion field $\Phi$,

\ep \[
P(U,\Phi) = \frac{1}{Z} e^{-[S_W(U)+\Phi^*
(\mathcal{M}^\dagger\mathcal{M})^{-1}\Phi]} = e^{-\seff}.
\]\ep

\noindent The staggered Dirac operator is $\mathcal{M}$ with the
pseudofermion field only defined on even sites.

The $\Phi$ algorithm \cite{ralg} is a Hybrid Molecular Dynamics (HMD)
algorithm for 4 flavours of staggered fermions.  It iterates a
composite Markov step, which is ergodic and has a fixed point
distribution close to the desired one.  We introduce conjugate momenta
$\pi$ in order to define a Hamiltonian $H$,

\ep \[
P(U,\Phi,\pi) = \frac{1}{Z'} e^{-[\frac{1}{2}\pi^2 + \seff]} =
\frac{1}{Z'}e^{-H(U,\Phi,\pi)}\;.
\]\ep

\noindent The action $\seff$ takes the role of the potential in the
Hamiltonian.  The gauge fields $U$ can then be allowed to evolve for a
time $\tau$ by integrating Hamilton's equations, using a Molecular
Dynamics (MD) integration scheme.  Each MD trajectory consists of a
momentum refreshment heatbath using Gaussian noise, a pseudofermion
heatbath using Gaussian noise, and finally an MD trajectory consisting
of $\tau/\delta\tau$ steps.

Typically a $QPQ$ symmetric symplectic integrator is used, that is one
which evolves $U$ by half a step, $\pi$ by a step and finally $U$ by a
half step.  This does not conserve energy, having $\delta
H=O(\delta\tau^2)$ for
\begin{itshape}any\end{itshape} trajectory length.

For $\delta\tau > 0$, the fixed point distribution of the MD step and
the momentum refreshment heatbath do not coincide.  We must find the
actual equilibrium distribution of the composite of these two steps.
Since we discard the new momenta and pseudofermions after each step we
consider the full Markov step as an update of $U$ alone, integrating
out the auxilliary fields $\pi$ and $\Phi$.  Let $V(\tau)$ represent
the evolution operator for the MD step $V(\tau):(U,\pi)\mapsto
(U'',\pi'')$ and $e^{-(S+\Delta S)}$ denote the fixed point
distribution of the full Markov step, where $\Delta S$ measures the
deviation from the desired distribution.  This must satisfy

\begin{eqnarray}
\lefteqn{e^{-[S(U')+\Delta S(U')]}} \nonumber \\
&=& \int dU\,d\pi\,e^{-H(U,\pi)-\Delta S(U)}\delta(U'-U'') \nonumber \\
&=& \int dU''\,d\pi''\,e^{-(H+\Delta S)\circ V^{-1}}\delta(U'-U'')\nonumber\\
&=& \int dU''\,d\pi''\,e^{-(H+\Delta S)} e^{-\delta(H+\Delta S)}\nonumber\\
&=& e^{-[S(U')+\Delta S(U')]} \langle e^{-\delta(H+\Delta S)}\rangle_\pi\nonumber,
\end{eqnarray}

\noindent with $\delta :\Omega\mapsto\Omega\circ[V(\tau)-1]$ measuring
the change in $\Omega$ over a trajectory (i.e., $\delta H$ is the
extent to which energy is not conserved).  We have assumed
reversibility $V^{-1}=F\circ V\circ F$ where $F:(U,\pi)\mapsto(U,-\pi)$
and area preservation so the Jacobian is unity.  Thus we obtain the
condition

\ep
\begin{equation}
\langle e^{-\delta(H+\Delta S)}\rangle_\pi = 1.
\label{condition}
\end{equation} \ep

\noindent Performing an asymptotic expansion on this condition in
powers of $\delta\tau$, knowing $\delta H=O(\delta\tau^2)$ for any
trajectory length $\tau$, we deduce that $\delta\Delta S\sim
O(\delta\tau^2)$.  We thus have $\Delta S \sim O(\delta\tau^2)$ for
$\tau \gg \delta\tau$, hence $\Phi$ is $O(\delta\tau^2)$ accurate.

\subsection{The $\chi$ Algorithm}
This is very similar to $\Phi$, except the pseudofermion heatbath is
performed before every single MD step as opposed to only before each
MD trajectory.  The proof that the leading error is $O(\delta\tau^2)$
follows from that of $R$ given later.

\subsection{The $R_0$ algorithm}
$R_0$ begins from a completely different viewpoint.  Instead of
introducing pseudofermions to replace the fermionic determinant, we
include the determinant in the action\cite{ralg}.  The fermionic
action is thus

\ep \[
S_F = -n\tr\ln\mathcal{M}^{\dagger} \mathcal{M},
\]\ep

\noindent where $n$ is the number of fermion multiplets ($n=N_f/4$ for
staggered fermions).  When computing the fermionic force contribution,
a noisy field estimator is used for the trace.

The ergodic composite Markov step now consists of a momentum
refreshment heatbath and an MD trajectory consisting of
$N\equiv\tau/\delta\tau$ steps with independent noise $\eta$ used for
each step.

To calculate the error in $\delta H$ we have to include the effect of
using a noisy estimator, $\langle\Sigma'\rangle_\eta=S'$ for the
fermionic force.  For a single noisy MD step we find that

\ep \[
\langle e^{-\delta H}\rangle_{\eta} =
\frac{1}{2}\langle(\Sigma'-S')^2\rangle_{\eta}(1-\pi^2)\delta\tau^2 +
O(\delta\tau^3).
\]\ep

\noindent The coefficient of $(1-\pi^2)\delta\tau^2$ is proportional to
the variance of the estimated force and will only vanish if the force
is computed exactly.  Since the momentum average is not Gaussian after
many leapfrog steps, the leading order term is not cancelled as it
would be if we only did one MD step per trajectory.  Thus over an
entire trajectory $\delta H\sim O(\delta\tau)$ and the leading error
of $R_0$ is $\Delta S\sim O(\delta\tau)$.

\subsection{The $R$ algorithm}

The only difference between $\chi$ and $R_0$ is that in the former the
pseudofermion field is calculated at the beginning of every MD step,
and in the latter the noisy field (effectively pseudofermions) is
calculated in the middle of each MD step.  $\chi$ has
$O(\delta\tau^2)$ errors for $n=1$ multiplets, whereas $R_0$ has
errors $O(\delta\tau)$.  However for $n=0$ (i.e., no fermions) both
are identical and have errors of $O(\delta\tau^2)$.  We expect that
the leading error has a linear dependence on the time the
pseudofermions are refreshed and on the number of multiplets, so if we
refresh the pseudofermion field at $t = (1-n)\delta\tau/2$, an
$O(\delta\tau^2)$ algorithm for $0\le n\le 1$ fermion multiplets
should be obtained.  For two flavours of staggered fermions, this
means evaluating the pseudofermion field a quarter way through each MD
update\cite{ralg}.  This leads to an algorithm that is neither
reversible nor area preserving and so cannot be made exact (unlike the
previous algorithms which could be made exact through the inclusion of
an accept/reject step).

The argument leading to Equation (\ref{condition}) may be generalised to give

\ep \begin{equation}
\langle e^{(\delta+\bar{\delta})(H+\Delta S)-\tr\ln\,V_*}\rangle_\pi=1,
\label{req}
\end{equation} \ep

\noindent where $\delta$ measures lack of ``energy'' conservation,
$\bar{\delta}$ measures lack of reversibility $\bar{\delta}:
\Omega\mapsto \Omega\circ [V(\tau)^{-1}-F\circ V\circ F]$ and
$\tr\ln\, V_*\equiv\ln\det\frac{\partial(U'',\pi'')} {\partial(U,\pi)}$
measures lack of area preservation.  Considering a single step of the
$R$ algorithm, where the auxilliary field $\chi$ is computed at a time
$t=(1-\alpha)\delta\tau/2$ where $\alpha$ is some parameter to be
determined, and expanding Equation (\ref{req}) in $\delta\tau$ we find

\ep \begin{equation}
\langle e^{-(\delta+\bar{\delta})(H+\Delta
S)-\tr\ln\,V_*}\rangle_{\pi}=1 - A\delta\tau^2 + O(\delta\tau^3),
\label{rexp}
\end{equation} \ep

\noindent where $A$ is proportional to $(n-\alpha)$.  If $\alpha=n$ the leading
term cancels, and thus the leading error is $O(\delta\tau^2)$ for the
entire trajectory.  Therefore, as claimed $R$ is an $O(\delta\tau^2)$
algorithm, and thus so is $\chi$ (i.e., $R$ with $n=1$).

\section{A Source of Inaccuracy}

The staggered fermion kernel is

\ep \[
\mathcal{M} = 2m\delta_{i,j} +
\sum_{\mu}\eta_{i,\mu}(U_{i,\mu}\delta_{i,j-\mu} -
U^{\dagger}_{i-\mu,\mu}\delta_{i,j+\mu}),
\]\ep

\noindent where $m$ is the fermion mass and $U_{i,\mu}$ is the gauge
field link matrix at site $i$ in direction $\mu$ and $\eta_{i,\mu}$
are the staggered fermion phase factors.  The $\delta\tau^3$ term in
Equation (\ref{rexp}) should have a coefficient that behaves as $\left
[(\mathcal{M}^{\dagger}\mathcal{M})^{-1} \frac{\partial}{\partial
U}(\mathcal{M}^{\dagger}\mathcal{M})\right ]^3$, thus for light modes
this term could be expected to behave as $O(m^{-3})$.  This presents
no problems, as long as $\delta \tau$ is small compared to the mass.
If $\delta\tau\approx m$, then the $\delta\tau$ expansion breaks down.
For an exact algorithm, the accept/reject step would have corrected
for this, however with an inexact algorithm $\Delta S \sim
O((\delta\tau/m)^2) = O(1)$, so we would be simulating an action
$S+\delta S$ which differs from $S$ by terms which are not small.

When short distance observables (e.g., the plaquette) are measured
with $\delta\tau\approx m$ (typical of light fermion simulations)
there is the expected $\delta\tau^2$ scaling, with no indication of
the inaccuracy just highlighted.  However the $m^{-3}$ behaviour would
only be expected to be true for the lightest fermion modes, and since
bosonic observables do not couple strongly to these modes; we do not
expect the $m^{-3}$ behaviour to be observable here.

Instantons correspond to zero modes of the Dirac operator in the
massless limit, and are crucial for physical dynamical quark effects,
such as the $\eta '$ mass.  An error $\Delta S \sim O(1)$ for the
lightest modes would thus most likely affect the instanton sector, and
thus one may get the most interesting light dynamical quark physics
wrong.  Unfortunately accurate measurement of such effects is
notoriously hard.

\section{The RHMC algorithm}
\ep
The RHMC is an exact algorithm which can be used to to simulate two
flavours of staggered fermions.  It is like HMC with two extra
ingredients: a fairly cheap but very accurate force computation and a
cheap noisy accept/reject step\cite{rhmc}\cite{jlqcd}.

We now write the fermion action as

\ep \[
S_F=\chi^{\dagger}(\mathcal{M}^{\dagger} \mathcal{M})^{-1/2} \chi.
\]\ep

\noindent The inverse square root of the Dirac operator, can be
approximated using an optimal Chebyshev rational approximation.  The
advantage of rational approximations is that the error in the
approximation falls as $e^{n/\ln\,m}$ where $n$ is the degree of the
rational function used and $m$ is the fermion mass.  This accuracy is
maintained over the entire spectrum of the Dirac operator.

The noisy part of the accept/reject step corrects the
errors in the approximation of the square-root, the $\delta\tau$
errors being corrected exactly.

\section{Testing the $R$ algorithm}

We propose to test $R$ by comparing it against RHMC.  Initially RHMC
will be used to generate thermalised ensembles which will then be
evolved using $R$.  Changes in observables and/or autocorrelation
lengths would signal $R$ does not get the correct distribution.

Initial work has begun on the testing, although it is still too early
to reach any conclusions.

\section{Acknowledgments}
Our thanks go to Robert Edwards, Ivan Horv\'{a}th, Stefan Sint and Urs
Wenger for discussions and for providing the RHMC code in the SZIN
software system\cite{szin}.

Work supported by the European Community's Human potential programme
under HPRN-CT-2000-00145 Hadrons/LatticeQCD.

\bibliography{proceedings}
\bibliographystyle{h-physrev}

\end{document}